\documentclass[sigconf]{acmart}
\usepackage{enumitem}
\usepackage{flushend}
\usepackage{xspace}
\usepackage{listings}
\usepackage{xcolor}
\usepackage{algorithm}
\usepackage{algorithmicx}
\usepackage{amsmath}
\usepackage{amsfonts}
\usepackage{graphicx}
\usepackage{algpseudocode}
\usepackage{amsthm}
\usepackage{soul}
\usepackage{hyperref}
\usepackage{tikz}
\usepackage{pgfplots}
\usepackage{booktabs}
\usepgfplotslibrary{statistics}
\usetikzlibrary{automata,arrows}
\newcommand{\RQ}[1]{\textbf{RQ#1}\xspace} % research questions

\newcommand{\roundbox}[1]{
\begin{center}
	\begin{tikzpicture}
		\node[draw=black, rectangle, rounded corners](box){
    	\begin{minipage}{\columnwidth}
        	#1
    	\end{minipage}
		};
	\end{tikzpicture}
\end{center}
}

\newcommand{\lineref}[1]{Line~\ref{#1}\xspace}
\newcommand{\figref}[1]{Figure~\ref{#1}\xspace}
\newcommand{\secref}[1]{Section~\ref{#1}\xspace}

\makeatletter
\renewcommand*{\NAT@spacechar}{~}
\makeatother

\newcommand{\empirical}[1]{#1}

\newcommand{\DSL}{\textsc{CrySL}\xspace}
\newcommand{\TOOLSA}{\textsc{CogniCrypt}$_{\textit{\textsc{sast}}}$\xspace}

\newcommand{\cryptolint}{\textsc{CryptoLint}\xspace}
\newcommand{\crapis}{Crypto~APIs\xspace}
\newcommand{\SARuleSet}{\textsc{Ruleset}$_{\textit{\textsc{sast}}}$\xspace}
\newcommand{\CLRules}{\textsc{Ruleset}$_{\textit{\textsc{cl}}}$\xspace}
\newcommand{\CLDSLRuleSet}{\textsc{Ruleset}$_{\textit{\textsc{cl\DSL}}}$\xspace}

\newcommand{\mathFont}[1]{\mathbb{#1}}
\newcommand{\code}[1]{\texttt{#1}}
   
\definecolor{javared}{rgb}{0.6,0,0} % for strings
\definecolor{javagreen}{rgb}{0.25,0.5,0.35} % comments
\definecolor{javapurple}{rgb}{0.5,0,0.35} % keywords
\definecolor{javadocblue}{rgb}{0.25,0.35,0.75} % javadoc   

\lstdefinestyle{CryptSL}{language=Java,
stringstyle=\color{blue},
keywords={SPEC, OBJECTS, FORBIDDEN, EVENTS, ORDER, CONSTRAINTS, ENSURES, REQUIRES, NEGATES}
}

\newcommand{\dslCommand}[1]{\textbf{\code{#1}}\xspace}
\newcommand{\NEGATES}{\dslCommand{NEGATES}}
\newcommand{\ENSURES}{\dslCommand{ENSURES}}
\newcommand{\REQUIRES}{\dslCommand{REQUIRES}}
\newcommand{\FORBIDDENEVENTS}{\dslCommand{FORBIDDEN}}
\newcommand{\EVENTS}{\dslCommand{EVENTS}}
\newcommand{\ORDER}{\dslCommand{ORDER}}
\newcommand{\CONSTRAINTS}{\dslCommand{CONSTRAINTS}}
\newcommand{\OBJECTS}{\dslCommand{OBJECTS}}
\newcommand{\SPEC}{\dslCommand{SPEC}}
\newcommand{\IDEAL}{IDE$^{al}$\xspace}

\newtheorem{defn}{Definition}

\newcommand{\objs}{\mathcal{O}}
\newcommand{\vals}{\mathcal{V}}

\newcommand{\methods}{\mathFont{M}}
\newcommand{\var}{\mathFont{V}}

\newcommand{\events}{\mathFont{E}}
\newcommand{\predicates}{\mathFont{P}}
\newcommand{\bool}{\mathFont{B}}
\newcommand{\cryptSLrule}{\mathFont{SPEC}}
\newcommand{\constraints}{\mathFont{C}}
\newcommand{\specType}{T}
\newcommand{\forbiddenEvents}{\mathcal{E}}
\newcommand{\usagePattern}{\mathcal{A}}
\newcommand{\constraint}{\mathcal{C}}
\newcommand{\genPreds}{\mathcal{G}}

\newcommand{\automatons}{\mathFont{A}}

\newcommand{\powerset}[1]{\mathcal{P}(#1)}
\newcommand{\sat}{sat}

\newif\ifnotfinal
\notfinaltrue
\definecolor{darkred}{rgb}{0.75,0,0}

\setcopyright{rightsretained}
\acmConference[ICSE'18]{International Conference on Software Engineering}{May-June 2018}{Gothenburg, Sweden} 
\acmYear{2018}
\copyrightyear{2017}

\begin{document}
\title{\DSL: Validating Correct Usage of Cryptographic APIs}

\author{Stefan Kr{\"u}ger}
\affiliation{Paderborn University}
\email{stefan.krueger@upb.de}

\author{Johannes Sp{\"a}th}
\affiliation{Fraunhofer IEM}
\email{johannes.spaeth@iem.fraunhofer.de}

\author{Karim Ali}
\affiliation{University of Alberta}
\email{karim.ali@ualberta.ca}

\author{Eric Bodden}
\affiliation{\mbox{Paderborn University \& Fraunhofer IEM}}
\email{eric.bodden@upb.de}

\author{Mira Mezini}
\affiliation{TU Darmstadt}
\email{mezini@cs.tu-darmstadt.de}

\begin{abstract}
Various studies have empirically shown that the majority of Java and Android apps misuse cryptographic libraries, causing devastating breaches of data security. Therefore, it is crucial to detect such misuses early in the development process. The fact that insecure usages are not the exception but the norm precludes approaches based on property inference and anomaly detection.

In this paper, we present \DSL, a definition language that enables cryptography experts to specify the secure usage of the cryptographic libraries that they provide. \DSL combines the generic concepts of method-call sequences and data-flow constraints with domain-specific constraints related to cryptographic algorithms and their parameters.
We have implemented a compiler that translates a \DSL ruleset into a context- and flow-sensitive demand-driven static analysis. The analysis automatically checks a given Java or Android app for violations of the \DSL-encoded rules.

We empirically evaluated our ruleset through analyzing \empirical{10,001} Android apps. Our results show that misuse of cryptographic APIs is still widespread, with \empirical{96\%} of apps containing at least one misuse. However, we observed fewer of the misuses that were reported in previous work.
\end{abstract}


\maketitle

\section{Introduction}\label{sec:intro}
Digital devices are increasingly storing sensitive data, which is often protected using cryptography. However, it is insufficient to use secure cryptographic algorithms. A developer must also know how to \emph{securely use} such algorithms in their code. Unfortunately, prior studies suggest that this is rarely the case. \citet{LazarCWZ14} examined \empirical{269} published cryptography-related vulnerabilities. They found that \empirical{223} are caused by developers misusing a security library, and only \empirical{46} result from faulty library implementations. \citet{EgeleBFK13} statically analyzed  \empirical{11,748} Android apps using cryptography-related application interfaces (\crapis) and found \empirical{88\%} of them violated at least one basic cryptography rule. \citet{Chatzikonstantinou:2016} reached a similar conclusion by first analyzing apps dynamically and then performing a manual inspection for misuses.

Such pervasive insecure use of cryptographic libraries is problematic for several reasons. First, the misuses of \crapis lead to devastating data breaches in a large number of applications. \citet{blackhat15baas} show that \emph{virtually all} smartphone apps that rely on cloud services use hard-coded keys. A simple decompilation gives adversaries access to those keys, and to all data that all these apps store in the cloud. \citet{NadiKMB16} were the first to investigate why developers often struggle to use \crapis. The authors conducted four studies, two of which survey Java developers familiar with the Java \crapis. The majority of participants (\empirical{65\%}) found their respective \crapis hard to use. When asked why, participants mentioned the API level of abstraction, insufficient documentation without examples, and API design make it difficult to understand how to properly use \crapis. A potential long-term solution is redesigning the APIs to provide an easy-to-use interface for developers that is secure by default. However, it remains crucial to detect and fix the existing insecure API uses. When asked about what would simplify their API usage, participants wished they had tools that help them automatically detect misuses and suggest possible fixes~\citet{NadiKMB16}. Unfortunately, approaches based solely on specification inference and anomaly detection~\cite{tse12automated} are not viable for \crapis, because---as elaborated above---most uses of \crapis are insecure.

In this paper, we present (a) \DSL, a definition language that enables cryptography experts to specify the secure usage of their \crapis, and (b) \TOOLSA, a compiler that parses and type-checks \DSL rules and translates them into a static analysis. The analysis automatically checks a given Java or Android app for compliance with the encoded \DSL rules. \DSL goes beyond methods that are useful for general validation of API usage (e.g., typestate analysis~\cite{AllanACHKLMSST05,BierhoffA07, NaeemL08, ICSE10Cont} and data-flow checks~\cite{AlhadidiBBDB09,ArztRFBBKTOM14}) by enabling the expression of domain-specific constraints related to cryptographic algorithms and their parameters. Our focus is the Java Cryptography Architecture (JCA), because it is the primary cryptography API for Java applications~\cite{NadiKMB16}. To evaluate \DSL, we encoded a comprehensive ruleset for the JCA classes and interfaces, and we used the generated static analysis to scan \empirical{10,001} Android apps that use the JCA. \TOOLSA found at least one misuse in \empirical{96\%} of the apps. For more than \empirical{75\%} of the apps, \TOOLSA finishes in under \empirical{4}~minutes.

In summary, this paper presents the following contributions:
\begin{itemize}
\item We introduce \DSL, a definition language to specify correct usages of \crapis.
\item We encode a comprehensive specification of correct usages of the JCA in \DSL.
\item We present \TOOLSA, a compiler that translates \DSL rules into a static analysis to find violations in a given Java or Android app.
\item We empirically evaluate \TOOLSA on \empirical{10,001} Android apps.
\end{itemize}

We will open source our implementation and artifacts on GitHub.
\section{Motivating Example}

\begin{figure}
\begin{lstlisting}
SecretKeyGenerator kG = KeyGenerator.getInstance("AES"); /*@\label{exKeyGenbeg}@*/
kG.init(128); /*@\label{exKeyGeninter}@*/
SecretKey cipherKey = kG.generateKey(); /*@\label{exKeyGenend}@*/

String plaintextMSG = getMessage(); /*@\label{exCipherGenbeg}@*/
Cipher ciph = Cipher.getInstance("AES/GCM"); /*@\label{exCipherGen}@*/
ciph.init(Cipher.ENCRYPT_MODE, cipherKey); /*@\label{exCipherGenend}@*/
byte[] cipherText = ciph.doFinal(plaintextMSG.getBytes("UTF-8"));
\end{lstlisting}
\caption{An example illustrating the use of \code{javax.crypto.KeyGenerator} to implement data encryption in Java.}
\label{lst:exKeygen-cipher}
\end{figure}
\begin{figure}
\footnotesize
\begin{algorithmic}
	\State
	\State \hspace{\algorithmicindent} METHOD := 
	\State \hspace{\algorithmicindent} \hspace{\algorithmicindent} methname(PARAMETERS)
	\State
	
	\State \hspace{\algorithmicindent} PARAMETERS := 
	\State \hspace{\algorithmicindent} \hspace{\algorithmicindent} varname \textbf{,} PARAMETERS 
	\State \hspace{\algorithmicindent} \hspace{\algorithmicindent} varname
	\State
	
	\State \hspace{\algorithmicindent} TYPES := 
	\State \hspace{\algorithmicindent} \hspace{\algorithmicindent} QualifiedClassName \textbf{,} TYPES 
	\State \hspace{\algorithmicindent} \hspace{\algorithmicindent} TYPE
	\State
	
	\State \hspace{\algorithmicindent} CONSTANTLIST := 
	\State \hspace{\algorithmicindent} \hspace{\algorithmicindent} constant \textbf{,} CONSTANTLIST 
	\State \hspace{\algorithmicindent} \hspace{\algorithmicindent} constant
	\State
	
	\State \hspace{\algorithmicindent} AGGREGATE := 
	\State \hspace{\algorithmicindent} \hspace{\algorithmicindent} label \textbf{$|$} AGGREGATE 
	\State \hspace{\algorithmicindent} \hspace{\algorithmicindent} label \textbf{;}
	\State
	
	\State \hspace{\algorithmicindent} EVENT := 
	\State \hspace{\algorithmicindent} \hspace{\algorithmicindent} AGGREGATE
	\State \hspace{\algorithmicindent} \hspace{\algorithmicindent} label \textbf{:} METHOD
	\State \hspace{\algorithmicindent} \hspace{\algorithmicindent} label \textbf{:} varname \textbf{=} METHOD  {\hfill \textit{A: B = C(D) --- a single event with}}
	\State {\hfill \textit{label A consisting of method C, its}}
	\State {\hfill \textit{parameter D, and return object B}}
	\State 
	
	\State \hspace{\algorithmicindent} PREDICATE := 
	\State \hspace{\algorithmicindent} \hspace{\algorithmicindent}  predname(PARAMETERS)	
	\State \hspace{\algorithmicindent} \hspace{\algorithmicindent}  predname(PARAMETERS) after EVENT
	
	\end{algorithmic}
\caption{Basic \DSL syntax elements.}
\label{def:syntax-basic}
\end{figure}
\begin{figure}
\footnotesize
\begin{algorithmic}
\State \textbf{SPEC} TYPE;
\State 

\State \textbf{OBJECTS}
	\State \hspace{\algorithmicindent} OBJECTS :=
	\State \hspace{\algorithmicindent} \hspace{\algorithmicindent} OBJECT \textbf{;} OBJECTS {\hfill \textit{A ; B --- a list of objects A and B}}
	\State \hspace{\algorithmicindent} \hspace{\algorithmicindent} OBJECT \textbf{;} {\hfill \textit{A --- a list of the single object A}}
	
	\State \hspace{\algorithmicindent} OBJECT := 
	\State \hspace{\algorithmicindent} \hspace{\algorithmicindent}	TYPE varname {\hfill \textit{A B --- object B of Java type A}}
	\State 
	
\State \textbf{EVENTS}	
	\State \hspace{\algorithmicindent} EVENTS :=
	\State \hspace{\algorithmicindent} \hspace{\algorithmicindent}	EVENT \textbf{;} EVENTS {\hfill \textit{A ; B --- a list of events A and B}}
	\State \hspace{\algorithmicindent} \hspace{\algorithmicindent}	EVENT \textbf{;} {\hfill \textit{A --- a list of the single event A}}
	\State
		
\State \textbf{FORBIDDEN}
	\State \hspace{\algorithmicindent} FMETHODS :=
	\State \hspace{\algorithmicindent} \hspace{\algorithmicindent}	FMETHOD \textbf{;} FMETHODS {\hfill \textit{A ; B --- a list of forbidden A and B}}
	\State \hspace{\algorithmicindent} \hspace{\algorithmicindent}  FMETHOD \textbf{;} {\hfill \textit{A --- a list of the single forbidden method A}}
	
	\State \hspace{\algorithmicindent} FMETHOD :=
	\State \hspace{\algorithmicindent} \hspace{\algorithmicindent}	methname(TYPES) \verb!=>! label {\hfill \textit{A(B)} \verb!=>! \textit{C --- a forbidden method
	named A}}
	\State {\hfill \textit{with parameter of Type B and replacement C}}

	\State
\State \textbf{ORDER}
	\State \hspace{\algorithmicindent} 
	USAGEPATTERN :=
	\State \hspace{\algorithmicindent} \hspace{\algorithmicindent} USAGEPATTERN \textbf{,} USAGEPATTERN {\hfill \textit{A , B --- A followed by B}}
	\State \hspace{\algorithmicindent} \hspace{\algorithmicindent} USAGEPATTERN $|$ USAGEPATTERN {\hfill \textit{A $|$ B --- A or B}} 
	\State \hspace{\algorithmicindent} \hspace{\algorithmicindent} USAGEPATTERN \textbf{?} {\hfill \textit{A? --- A is optional}} 
	\State \hspace{\algorithmicindent} \hspace{\algorithmicindent} USAGEPATTERN \textbf{*} {\hfill \textit{A* --- 0 or more As}} 
	\State \hspace{\algorithmicindent} \hspace{\algorithmicindent} USAGEPATTERN \textbf{+} {\hfill \textit{A+ --- 1 or more As}} 
	\State \hspace{\algorithmicindent} \hspace{\algorithmicindent} \textbf{(} USAGEPATTERN \textbf{)} {\hfill \textit{(A) --- grouping}}  
	\State \hspace{\algorithmicindent} \hspace{\algorithmicindent} AGGREGATE 
	\State

\State \textbf{CONSTRAINTS}
	\State \hspace{\algorithmicindent} CONSTRAINTS := 
	\State \hspace{\algorithmicindent} \hspace{\algorithmicindent} CONSTRAINT \textbf{,} CONSTRAINTS 
	\State \hspace{\algorithmicindent} \hspace{\algorithmicindent} CONSTRAINT
	\State \hspace{\algorithmicindent} CONSTRAINT := 
	\State \hspace{\algorithmicindent} \hspace{\algorithmicindent} varname \textbf{in} \textbf{\{} CONSTANTLIST \textbf{\}}  {\hfill \textit{A in \{1, 2\} --- A should be 1 or 2}} 
	\State
	
	\State \textbf{REQUIRES}
	\State \hspace{\algorithmicindent} REQ\_PREDICATES := 
	\State \hspace{\algorithmicindent} \hspace{\algorithmicindent} PREDICATE
	\State
	
	\State \textbf{ENSURES}
	\State \hspace{\algorithmicindent} ENS\_PREDICATES := 
	\State \hspace{\algorithmicindent} \hspace{\algorithmicindent} PREDICATE
	\State
	
	\State \textbf{NEGATES}
	\State \hspace{\algorithmicindent} NEG\_PREDICATES := 
	\State \hspace{\algorithmicindent} \hspace{\algorithmicindent} PREDICATE
\end{algorithmic}
\caption{A \DSL rule in Extended Backus-Naur Form (EBNF)~\cite{DBLP:journals/cacm/BackusBGKMPRSVWWWN63}.}
\label{def:syntax}
\end{figure}

Throughout the paper, we will use the code example in \figref{lst:exKeygen-cipher} to motivate the language features in \DSL. Lines~\ref{exKeyGenbeg}--\ref{exKeyGenend} generate a 128-bit secret key to use with the encryption algorithm AES. Lines~\ref{exCipherGenbeg}--\ref{exCipherGenend} use that key to initialize a Java \code{Cipher} object that encrypts \code{plaintextMSG}. Since AES encrypts plaintext block by block, it must be configured to use one of several \emph{modes of operation}. The mode of operation determines how to encrypt a block based on the encryption of the preceding block(s). \lineref{exCipherGen} configures \code{Cipher} to use the Galois/Counter Mode (\code{GCM}) of operation~\cite{McGrewV04}.

Although the code example may look straightforward, there are a number of subtle mistakes that render the encryption insecure. First, both \code{KeyGenerator} and \code{Cipher} only support a limited choice of encryption algorithms. If the developer passes an unsupported algorithm to either \code{getInstance} methods, the respective line will throw a runtime exception. Similarly, the design of the APIs separates the classes for key generation and encryption. Therefore, the developer needs to make sure they pass the same algorithm to the \code{getInstance} methods of \code{KeyGenerator} and \code{Cipher}. If the developer does not configure the algorithms as such, the generated key will not fit the encryption algorithm, and the encryption will fail by throwing a runtime exception. Moreover, some supported algorithms are no longer considered secure (e.g., DES or AES/ECB~\cite{BSI}). If the developer selects such an algorithm, the program will still run to completion, but the resulting encryption will be insecure.

To use \crapis properly, developers have to take two dimensions of correctness into consideration: (1) the functional correctness that allows the program to run and terminate successfully and (2) the provided security guarantees. Prior empirical studies have shown that developers frequently succeed in obtaining functional correctness by, for instance, looking for code examples on web portals such as StackOverflow~\cite{FischerBXSA0F17}. However, they often fail to obtain a secure use of \crapis, primarily because most code examples on those web portals provide insecure solutions~\cite{FischerBXSA0F17}.

\section{\DSL Syntax}
\label{sec:lang}

Instead of relying on the security of existing usages and examples, we present an approach in which cryptography experts define correct API usages in a domain-specific language, \DSL. In this section, we give an overview of the \DSL syntax elements. 
A formal treatment of the \DSL semantics is presented in \secref{syntaxSemantics}. \figref{def:syntax-basic} presents the basic syntactic elements of \DSL, and \figref{def:syntax} presents the full syntax for a \DSL rule. \figref{lst:exCryptSLKeyGen} shows an abbreviated \DSL rule that defines the correct usage of \code{javax.crypto.KeyGenerator} in the example in \figref{lst:exKeygen-cipher}.

\subsection{Mandatory Sections in a \DSL Rule}
To provide simple and reusable constructs, a \DSL rule is defined on the level of individual classes. Therefore, the rule starts off by stating the class that it is defined for. 

In \figref{lst:exCryptSLKeyGen}, the \OBJECTS section defines four objects to be used in later sections of the rule (e.g., the object \code{algorithm} of type \code{String}). These objects are typically used as parameters or return values in the \EVENTS section.

The \EVENTS section defines all methods that may contribute to the successful usage of a \code{KeyGenerator} object, including three \code{getInstance} methods that are defined by two \emph{method event patterns} (Lines~\ref{exCryptSLKeyGen-g1}--\ref{exCryptSLKeyGen-g2}). The first parameter of all three methods is a \code{String} object whose value states the algorithm that the key should be generated for. This parameter is represented by the previously defined \code{algorithm} object. Two of the \code{getInstance} methods are overloaded with two parameters. Since we do not need to specify the second parameter in either method, we substitute it with an underscore that serves as a placeholder in one combined pattern definition (\lineref{exCryptSLKeyGen-g2}). Finally, the rule defines patterns for the various \code{init} methods that set the proper parameter values (e.g., keysize) and a \code{generateKey} method that completes the key generation and returns the generated key.

\lineref{exCryptSLKeyGen-Order} defines a usage pattern for \code{KeyGenerator} using the keyword \ORDER. The usage pattern is a regular expression of method event patterns that are defined in \EVENTS. Although each method pattern defines a label to simplify referencing related events (e.g., \code{g1}, \code{i2}, and \code{GenKey}), it is tedious and error-prone to require listing all those labels again in the \ORDER section. Therefore, \DSL allows defining \emph{aggregates}. An aggregate represents a disjunction of multiple patterns by means of their labels. \lineref{exCryptSLKeyGen-gIAgg} defines an aggregate that groups the two \code{getInstance} patterns. Using aggregates, the usage pattern for \code{KeyGenerator} reads: there must be exactly one call to one of the \code{getInstance} methods, followed by an optional call to one of the \code{init} methods, and finally a call to \code{generateKey}. 

\begin{figure}
\begin{lstlisting}[style=CryptSL]
SPEC javax.crypto.KeyGenerator

OBJECTS
  java.lang.String algorithm;
  int keySize;
  javax.crypto.SecretKey key;

EVENTS
  g1: getInstance(algorithm); /*@\label{exCryptSLKeyGen-g1}@*/
  g2: getInstance(algorithm, _); /*@\label{exCryptSLKeyGen-g2}@*/
  GetInstance := g1 | g2; /*@\label{exCryptSLKeyGen-gIAgg}@*/

  i1: init(keySize);
  i2: init(keySize, _);
  i3: init(_);
  i4: init(_, _);
  Init := i1 | i2 | i3 | i4;
  
  GenKey: key = generateKey();

ORDER
  GetInstance, Init?, GenKey /*@\label{exCryptSLKeyGen-Order}@*/

CONSTRAINTS
  algorithm in {"AES", "Blowfish"}; /*@\label{exCryptSLKeyGen-Constraintsbeg}@*/
  algorithm in {"AES"} => keySize in {128, 192, 256};
  algorithm in {"Blowfish"} => keySize in {128, 192, 256, 320, 384, 448};/*@\label{exCryptSLKeyGen-Constraintsend}@*/
  
ENSURES 
  generatedKey[key, algorithm];
\end{lstlisting}
\caption{\DSL rule for using \code{javax.crypto.KeyGenerator}.}
\label{lst:exCryptSLKeyGen}
\end{figure}
\begin{figure}
\begin{lstlisting}[style=CryptSL]
SPEC javax.crypto.Cipher

OBJECTS
  int encmode;
  java.security.Key key;
  java.lang.String transformation;
  ...

EVENTS
  g1: getInstance(transformation); /*@\label{line:alg}@*/
  ...
  i1: init(encmode, key);

...

REQUIRES  
  generatedKey[key, alg(transformation)]; /*@\label{line:genkey}@*/

ENSURES
  encrypted[cipherText, plainText];
\end{lstlisting}
\caption{\DSL rule for using \code{javax.crypto.Cipher}.}
\label{lst:exCryptSLCipher}
\end{figure}

Following the keyword \CONSTRAINTS, Lines~\ref{exCryptSLKeyGen-Constraintsbeg}--\ref{exCryptSLKeyGen-Constraintsend} define the constraints for objects defined under \OBJECTS and used as parameters or return values in the \EVENTS section. In the abbreviated \DSL rule in \figref{lst:exCryptSLKeyGen}, the first constraint limits the value of \code{algorithm} to AES or Blowfish. For each algorithm, there is one constraint that restricts the possible values of \code{keysize}.

The \ENSURES section is the final mandatory construct in a \DSL rule. The section specifies predicates to govern interactions between different classes. For example, a \code{Cipher} object uses a key obtained from a \code{KeyGenerator}. 
The \ENSURES section specifies what a class provides, presuming that the object is used properly. For example, the \code{KeyGenerator} \DSL rule in \figref{lst:exCryptSLKeyGen} ends with the definition of a \emph{predicate} \code{generatedKey} with the generated key object and its corresponding algorithm as parameters. This predicate may be \emph{required} by the rule for \code{Cipher} or other classes that make use of such a key through the optional element of the \REQUIRES block as illustrated in Figure \ref{lst:exCryptSLCipher}.

We have enriched \DSL with several auxiliary functions. For example, in Figure \ref{lst:exCryptSLCipher}, the function \code{alg} extracts the encryption algorithm from \code{transformation} (\lineref{line:genkey}). This function is necessary, because \code{generatedKey} expects only the encryption algorithm as its second parameter, but \code{transformation} holds more values than just the algorithm (e.g., \lineref{exCipherGen} in \figref{lst:exKeygen-cipher}). 

\subsection{Optional Sections in a \DSL Rule}

A \DSL rule may contain optional sections that which we will showcase through the \DSL rule for \code{PBEKeySpec}. In \figref{lst:exCryptSLPBEKeySpec}, the \FORBIDDENEVENTS section specifies methods that should not be called, because calling them is always insecure. \code{PBEKeySpec} derives cryptographic keys from a user-given password. For security reasons, it is recommended to use a cryptographic salt for this operation. However, the constructor \code{PBEKeySpec(char[] password)} does not allow for a salt to be passed and the implementation in the default provider does not generate one. Therefore, this constructor should not be called, and any call to it should be flagged. Consequently, the \DSL rule for \code{PBEKeySpec} lists it in the \FORBIDDENEVENTS section (\lineref{exPBEKeySpec-f1}). In the case of \code{PBEKeySpec}, there is an alternative secure constructor (\lineref{exPBEKeySpec-c1}). \DSL allows specifying an alternative method pattern event using the arrow notation (\lineref{exPBEKeySpec-f1}).

In general, predicates are generated for a particular usage only if it follows the usage pattern defined in the \ORDER section and fulfils all constraints in the \CONSTRAINTS section of its corresponding rule. However, \code{PBEKeySpec} differs from that pattern. The class contains a constructor that receives a user-given password, but the method \code{clearPassword} deletes that password later. Consequently, a \code{PBEKeySpec} object fulfils its role after calling the constructor until \code{clearPassword} is called. To support this usage, \DSL allows specifying a method event pattern that if called, a predicate is generated using the keyword \dslCommand{after} (\lineref{exPBEKeySpec-ens}). \DSL further supports killing an existing predicate in the \NEGATES section (\lineref{exPBEKeySpec-neg}). 

\begin{figure}
\begin{lstlisting}[style=CryptSL]
SPEC javax.crypto.spec.PBEKeySpec

OBJECTS 
  char[] pw;
  byte[] salt;
  int it;
  int keylength; 	

EVENTS
  c1: PBEKeySpec(pw, salt, it, keylength); /*@\label{exPBEKeySpec-c1}@*/
  cP: clearPassword(); /*@\label{exPBEKeySpec-cP}@*/

FORBIDDEN
  PBEKeySpec(char[]) => c1; /*@\label{exPBEKeySpec-f1}@*/
  PBEKeySpec(char[],byte[],int) => c1;

ORDER
  c1,  cP
...

ENSURES
  keyspec[this, keylength] after c1; /*@\label{exPBEKeySpec-ens}@*/

NEGATES
  keyspec[this, _]; /*@\label{exPBEKeySpec-neg}@*/
\end{lstlisting}
\caption{\DSL rule for \code{javax.crypto.spec.PBEKeySpec}.}
\label{lst:exCryptSLPBEKeySpec}
\end{figure}

%Each rule defines its own set of predicates. So far, the authors have defined all predicates, both their names and corresponding parameters. In the future, when external cryptography experts begin specifying their own components in \DSL, we will provide support for defining new predicates.
\section{\DSL Formal Semantics}
\label{syntaxSemantics}
%In this section, we present the formal semantics of \DSL rules. 

\subsection{Basic Definitions}
A \DSL rule consists of several sections. 
The \OBJECTS section comprises a set of typed variable declarations $\var$. 
In the syntax in \figref{def:syntax}, each declaration $v\in \var$ is represented by the syntax element \code{TYPE varname}. The \EVENTS section contains elements of the form $(m,v)$, where $m\in \methods$ and $v \in \var^*$. $\methods$ is the set of all resolved method signatures, where each signature includes the method name and argument types. The \FORBIDDENEVENTS section lists a set of methods $\methods$ denoted by their signatures; forbidden events cannot bind any variables. The \ORDER section specifies the usage pattern in terms of a regular expression of labels or aggregates that are in $\methods$. We express this regular expression formally by the equivalent non-deterministic finite automaton $(Q, \methods, \delta, s_0, F)$ over the alphabet $\methods$, where $Q$ is a set of states, $q_0$ is its initial state, $F$ is the set of accepting states, and $\delta: Q \times \methods \rightarrow \powerset{Q}$ is the state transition function. 

The \CONSTRAINTS section is a subset of $\constraints := (\var \rightarrow \objs \cup \vals) \rightarrow \bool$, i.e., each constraint is a boolean function, where the argument is itself a function that maps variable names in $\var$ to objects in $\objs$ or values with primitive types in $\vals$. 

A \DSL rule is a tuple $(\specType, \forbiddenEvents, \usagePattern, \constraint)$, where $\specType$ is the reference type specified by the \SPEC keyword, $\forbiddenEvents \subseteq \methods$ is the set of forbidden events, $\usagePattern = (Q, \methods, \delta, s_0, F)  \in \automatons$ is the automaton induced by the regular expression of the \ORDER section, and $\constraint \subseteq \constraints$ is the set of \CONSTRAINTS that the rule lists. 

Our formal definition of a \DSL rule does not contain the sections \REQUIRES, \ENSURES, and \NEGATES. Those sections reason about the interaction of predicates, which requires a different formal treatment that we discuss in \secref{sec:predicates}.

\subsection{Runtime Semantics}\label{sub:semantics}

Each \DSL rule encodes usage constraints to be validated for all runtime objects of the reference type $\specType$ stated in its \SPEC section. We define the semantics of a \DSL rule in terms of an evaluation over a runtime program trace that records all relevant runtime objects and values, as well as all events specified within the rule.

\begin{defn}[Event]
An \emph{event} is a tuple $(m,e) \in \events$ of a method signature $m \in \methods$ and an \emph{environment} $e$, i.e., a mapping $\var \to \objs \cup \vals$ of the parameter variable names to concrete runtime objects and values. If the environment $e$ holds a concrete object for the \code{this} value, then it is called the event's \emph{base object}.
\end{defn}

\begin{defn}[Runtime Trace]\label{def:trace}
A \emph{runtime trace} $\tau \in \events^*$ is a finite sequence of events $\tau_0 \ldots \tau_n$.
\end{defn}

\begin{defn}[Object Trace]\label{def:obj-trace}
For any $\tau \in \events^*$, a subsequence $\tau_{i_1} ... \tau_{i_n}$ is called an \emph{object trace} if $i_1 < ... < i_n$ and all base objects of $\tau_{i_j}$ are identical. 
\end{defn}

Lines~\ref{exKeyGenbeg}--\ref{exKeyGeninter} in \figref{lst:exKeygen-cipher} result in an object trace that has two events:
\begin{align*}
&(m_0, \{algorithm \mapsto \code{"AES"}, \texttt{this}\mapsto o_{kg} \})\\
&(m_1, \{algorithm \mapsto \code{"AES"}, keySize \mapsto \code{128}, \texttt{this}\mapsto o_{kg} \})
\end{align*} 
where $m_0$ and $m_1$ are the signatures of the \code{getInstance} and \code{init} methods of the \code{KeyGenerator} class. 
For static factory methods such as \code{getInstance}, we assume \code{this} to bind to the returned object. $o_{kg}$ denotes the object that at runtime is bound to the variable \code{kG}.

The decision whether a runtime trace $\tau$ satisfies a set of \DSL rules involves two steps. In the first step, individual object traces are evaluated independently of one another. Yet, different runtime objects may still interact with each other. \DSL rules capture this interaction by means of predicates that a rule ensures on a runtime object. These interactions between different objects are checked in a second step against the specification by considering the predicates they require and ensure. We now discuss these steps in more detail.

\subsubsection{Individual Object Traces}

The sections \FORBIDDENEVENTS, \ORDER and \CONSTRAINTS are evaluated on individual object traces. \figref{fig:formalism} defines the function $\sat^o$ that is true if and only if a given trace $\tau^o$ for an object $o$ satisfies its \DSL rule. This definition of $\sat^o$ ignores interactions with other object traces. We will discuss later how such interactions are resolved.
In the following, we assume the trace $\tau^o = \tau^o_0, ..., \tau^o_n$, where $\tau^o_i=(m^o_i,e^o_i)$. We will also refer to our example from Figure \ref{lst:exKeygen-cipher} and the involved rules of \code{KeyGenerator} (\figref{lst:exCryptSLKeyGen}) and \code{Cipher} (\figref{lst:exCryptSLCipher}) to illustrate the computation.  The function $\sat^o$ is composed of three sub-functions:

\begin{figure}
\begin{align*}
\sat^o\colon \events^* \times \cryptSLrule  &\rightarrow \bool \\	
[\tau^o,(\specType^o,\forbiddenEvents^o, \usagePattern^o, \constraint^o)] \mapsto &\quad \sat^o_{F}(\tau^o,\forbiddenEvents^o) \\
&\wedge \sat^o_{\automatons}(\tau^o,\usagePattern^o)\\
&\wedge \sat^o_{\constraints}(\tau^o,\constraint^o) 
\end{align*}
\caption{The function $\sat^o$ verifies an individual object trace for the object $o$.}
\label{fig:formalism}
\end{figure}

\paragraph{Forbidden Events ($\sat^o_{F}$)}
Given a trace $\tau^o$ and a set of forbidden events $\forbiddenEvents$, $\sat^o$ ensures that none of the trace events is forbidden.
$$
\sat^o_{F}(\tau^o,\forbiddenEvents^o) := \bigwedge_{i=0\ldots n} m^o_i\notin \forbiddenEvents^o
$$

The \DSL rule for \code{KeyGenerator} does not list any forbidden methods. Hence, $\sat^o$ trivially evaluates to true for object \code{kG} in Figure \ref{lst:exKeygen-cipher}.

\paragraph{Order Errors ($\sat^o_{\automatons}$)}
The second function checks that the trace object is used in compliance with the specified usage pattern, i.e., all methods in the rule are invoked in no other than the specified order. Formally, the sequence of method signatures of the object trace $m^o := m^o_0, \ldots, m^o_n$, i.e., the projection onto the method signatures, must be an element of the language $\mathcal{L}(\usagePattern^o)$ that the automaton $\usagePattern^o = (Q, \methods, \delta, s_0, F)$ of the \ORDER section induces. By definition of language containment, after the last observed signature of the trace $m^o_n$, the corresponding state of the automaton must be an accepting state $s \in F$. This definition ignores any variable bindings. They are evaluated in the second step.
$$
\sat^o_{\automatons}(\tau^o,\usagePattern^o) :=  m^o \in \mathcal{L}(\usagePattern^o)
$$
\begin{figure}
	\centering
	\begin{tikzpicture}[->,>=stealth',shorten >=1pt,auto,node distance=1.9cm,
                    semithick]

  \node[state,initial]		(A)              {-1};
  \node[state] 				(B) [right of=A, node distance=2.6cm] {0};
  \node[state]      		(C) [right of=B] {1};
  \node[accepting,state]    (D) [right of=C] {2};

  \path (A) edge               node[below]      {\code{GetInstance}} (B)
  		(B) edge               node[below] {\code{Init}} (C)
        (C) edge       		   node[below] {\code{GenKey}} (D)
        (B) edge  [bend left]  node[above] {\code{GenKey}} (D);
\end{tikzpicture}
	\caption{The state machine for the \DSL rule in \figref{lst:exCryptSLKeyGen} (without an implicit error state).}
	\label{fig:smgKeyGen}
\end{figure}
Figure~\ref{fig:smgKeyGen} displays the automaton created for \code{KeyGenerator} using the aggregate names as labels. State \emph{-1} is the initial state, and state \emph{2} is the only accepting state. Following the code in \figref{lst:exKeygen-cipher} for the object \code{kG} of type \code{KeyGenerator}, the automaton transitions from state \emph{-1} to \emph{0} at the call to \code{getInstance} (\lineref{exKeyGenbeg}). With the calls to \code{init} (\lineref{exKeyGeninter}) and \code{generateKey} (\lineref{exKeyGenend}), the automaton first moves to state \emph{1} and finally to state \emph{2}. Therefore, function $
\sat^o_{\automatons}$ evaluates to true for this example.

\paragraph{Constraints ($\sat^o_{\constraints}$)}
The validity check of the constraints ensures that all constraints of $\constraint$ are satisfied. This check requires the sequence of environments $(e^o_0, ...,e^o_n)$ of the trace $\tau^o$. All objects that are bound to the variables along the trace must satisfy the constraints of the rule.
$$
\sat^o_{\constraints}(\tau^o,\constraint^o) := \bigwedge_{c \in \constraint^o, i=0\ldots n} c(e^o_i)
$$

To compute $\sat^o_{\constraints}$ for the \code{KeyGenerator} object \code{kG} at the call to \code{getInstance} in \lineref{exKeyGenbeg}, only the first constraint has to be checked. This is because the corresponding environment $e^o_{\ref{exKeyGenbeg}}$ holds a value only for \code{algorithm}, and the other two constraints reference other variable names. The evaluation function $c$ returns true if \code{algorithm} assumes either ``AES'' or ``Blowfish'' as its value, which is the case in \figref{lst:exKeygen-cipher}. The computation of $\sat^o_{\constraints}$ for Lines~\ref{exKeyGeninter}--\ref{exKeyGenend} works similarly.

\subsubsection{Interaction of Object Traces}\label{sec:predicates}
To define interactions between individual object traces, the \REQUIRES, \ENSURES, and \NEGATES sections allow individual \DSL rules to reference one another. For a rule for one object to hold at any given point in an execution trace, all predicates that its \REQUIRES section lists must have been both previously \emph{ensured} (by other specifications) and not \emph{negated}. Predicates are \emph{ensured} (i.e., generated) and \emph{negated} (i.e., killed) by certain events. Formally, a predicate is an element of $\predicates := \{(\textit{name}, \textit{args}) \mid \textit{args} \in \var^*\}$, i.e., a pair of a predicate name and a sequence of variable names. Predicates are generated in specific states. Each \DSL rule induces a function $\genPreds\colon S \rightarrow \powerset{\predicates}$ that maps each state of its automaton to the predicate(s) that the state generates.

The predicates listed in the \ENSURES and \NEGATES sections may be followed by the term \dslCommand{after n}, where $n$ is a method event pattern label or an aggregate. The states that follow the event or aggregate $n$ in the automaton generate the respective predicate. If the term \dslCommand{after} is not used for a predicate, the final states of the automaton generate (or negate) that predicate, i.e., we implicitly interpret it as \dslCommand{after n}, where $n$ is an event that leads to a final state. 

In addition to states that are selected as predicate-generating, the predicate is also ensured if the object resides in any state that transitively follows the selected state, unless the states are explicitly (de-)selected for the same predicate within the \NEGATES section.

At any state that generates a predicate, the event driving the automaton into this state binds the variable names to the values that the specification previously collected along its object trace.

Formally, an event $n^o=(m^o,e^o) \in \events$ of a rule $r$ and for an object $o$ ensures a predicate $p = (\textit{predName},\textit{args})\in \predicates$ on the objects $e^o\in \objs$ if: 

\begin{enumerate}
\item The method $m^o$ of the event leads to a state $s$ of the automaton that generates the predicate $p$, i.e. $p \in \genPreds(s)$. 
\item The runtime trace of the event's base object $o$ satisfies the function $\sat^o$.
\item All relevant \REQUIRES predicates of the rule are satisfied at execution of event $n^o$.
\end{enumerate}

For the \code{KeyGeneraor} object \code{kG} in Figure \ref{lst:exKeygen-cipher}, a predicate is generated at \lineref{exCipherGenend} because (1) its automaton transitions to its only predicate-generating state, state \emph{2}, (2) $\sat^o$ evaluates to true as previously shown for each subfunction and (3) the corresponding \DSL rule does not require any predicates.
\section{Detecting Misuses of \crapis}\label{sec:static-analysis}

To detect all possible rule violations, our tool \TOOLSA approximates the evaluation function $sat^o$ using a static data-flow analysis. In a security context, it is a requirement to detect as many misuses as possible. A drawback of this decision is the potential for false warnings that originate from over-approximations that the static analysis requires. In the following, we use the example in \figref{lst:MessageDigest} to illustrate why and where approximations are required. We will show later in our evaluation that, in practice, our analysis is highly precise and that the chosen approximations rarely lead to false warnings.

The code example in \figref{lst:MessageDigest} implements a hashing operation. By default, the code uses \code{SHA-256} for the operation. However, if the condition \code{option1} evaluates to true, \code{MD5} is chosen instead (\lineref{md-md5}). The \DSL rule for \code{MessageDigest}, displayed in \figref{lst:exCryptSLMessageDigest}, does not allow the usage of \code{MD5} though, because \code{MD5} is no longer secure~\cite{BSI}.

\begin{figure}

\begin{lstlisting}
boolean option1 = isPrime(66); //some non-trivial predicate returning false
byte[] input = "Message".getBytes("UTF-8");	

String alg = "SHA-256";
if (option1) alg = "MD5"; /*@\label{md-md5}@*/
MessageDigest md = MessageDigest.getInstance(alg);
	
if (input.size() > 0) md.update(input); /*@\label{branch}@*/
byte[] digest = md.digest();
\end{lstlisting}
\caption{An example illustrating the usage of \code{java.security.MessageDigest} in Java.}
\label{lst:MessageDigest}
\end{figure}

The \code{update} operation is performed only on non-empty input (\lineref{branch}). Otherwise, the call to \code{update} is skipped and only the call to \code{digest} is executed, without any input. Although not strictly insecure, this usage does not comply with the \DSL rule for \code{MessageDigest}, because it leads to no content being hashed.

\begin{figure}
\begin{lstlisting}[style=CryptSL]
SPEC java.security.MessageDigest

OBJECTS
  java.lang.String algorithm;
  byte[] input;
  int offset;
  int length;
  byte[] hash;
  ...

EVENTS
  g1: getInstance(algorithm);
  g2: getInstance(algorithm, _);
  Gets := g1 | g2;
  ...
  Updates := ...;

  d1: output = digest();
  d2: output = digest(input);
  d3: digest(hash, offset, length);
  Digests := d1 | d2 | d3;
  
  r: reset();

ORDER
  Gets, (d2 | (Updates+, Digests)), (r, (d2 | (Updates+, Digests)))*

CONSTRAINTS
  algorithm in {"SHA-256", "SHA-384", "SHA-512"};  

ENSURES
  digested[hash, ...];
  digested[hash, input];
\end{lstlisting}
\caption{\DSL rule for \code{java.security.MessageDigest}.}
\label{lst:exCryptSLMessageDigest}
\end{figure}

To approximate $\sat^o_{F}$, the analysis must search for possible forbidden events by first constructing a call graph for the whole program under analysis. It then iterates through the graph edges to find calls to forbidden methods. Depending on the precision of the call graph, the analysis may find calls to forbidden methods that cannot be reached at runtime.

The analysis represents each runtime object $o$ by its allocation site. In our example, allocation sites are \code{new} expressions and \code{getInstance} calls that return an object of a type for which a \DSL rule exists. For each such allocation site, the analysis approximates $\sat^o_{\automatons}$ by first creating a state-machine. \TOOLSA then evaluates the state machine using a typestate analysis that abstracts runtime traces by program paths. The typestate analysis is path-insensitive, thus, at branch points, it assumes that both sides of the branch may execute. In our example, this feature leads to a false positive: although the condition in \lineref{branch} always evaluates to true, and the call to \code{update} is never actually skipped, the analysis considers that this may happen, and thus reports a rule violation.

To approximate $\sat^o_{\constraints}$, we have extended the typestate analysis to also collect potential runtime values of variables along all program paths where an allocated object is used. The constraint solver first filters out all \emph{irrelevant} constraints. A constraint is irrelevant if it refers to one or more variables that the typestate analysis has not encountered. In \figref{lst:exCryptSLMessageDigest}, the rule only includes one internal constraint---on variable \code{algorithm}. If we add a new internal constraint to the rule about the variable \code{offset}, the constraint solver will filter it out as irrelevant when analyzing the code in \figref{lst:MessageDigest}, because the only method that this variable is associated with (\code{digest} labeled \code{d3}) is never called. The analysis distinguishes between never encountering a variable in the source code and not being able to extract the values of a variable. Using the same rule and code snippet, if the analysis fails to extract the value for \code{algorithm}, the constraint evaluates to false. Collecting potential values of a variable over all possible program paths of an allocation site may lead to further imprecision. In our example, the analysis cannot statically rule out that \code{algorithm} may be \code{MD5}. The rule forbids the usage of \code{MD5}. Therefore, the analysis reports a misuse.

Handling predicates in our analysis follows the formal description very closely. If $\sat^o$ evaluates to true for a given allocation site, the analysis checks whether all required predicates for the allocation site have been ensured earlier in the program. In the trivial case, when no predicate is required, the analysis immediately ensures the predicate defined in the \ENSURES section. The analysis constantly maintains a list of all ensured predicates, including the statements in the program that a given predicate can be ensured for. If the allocation site under analysis requires predicates from other allocation sites, the analysis consults the list of ensured predicates and checks whether the required predicate, with matching names and arguments, exists at the given statement. If the analysis finds all required predicates, it ensures the predicate(s) specified in the \ENSURES section of the rule.

% Due to the hierarchical and non-circular nature of dependencies between the predicates (i.e., a \code{Cipher} may depend on a \code{KeyGenerator} which in turn may depend on a \code{SecureRandom} object, but not the other way around), \REQUIRES blocks may not have circular dependencies. As typical in static-analysis implementations, \TOOLSA uses a worklist algorithm to resolve predicates in a fixed-point iteration. All rules are satisfied when in the fixed-point all required predicates have been ensured.
% \label{sec:worklist} 
\section{Implementation}\label{sec:implementation}

We have implemented \TOOLSA using Xtext~\cite{Xtext2017}, an open-source framework for developing domain-specific languages. Given the \DSL grammar, Xtext provides
a parser, type checker, and syntax highlighter for the language. When supplied with a type-safe \DSL rule, Xtext outputs the corresponding AST, which \TOOLSA then uses to generate the required static analysis. For the static analysis, we use the program analysis framework Soot~\cite{Vallee-Rai2000}. Soot transforms a given Java program into an intermediate representation that facilitates executing intra- and inter-procedural static analyses.
The framework provides standard static analyses such as call graph construction. 
Additionally, Soot can analyze a given Android app intra-procedurally. Further extensions by FlowDroid~\cite{ArztRFBBKTOM14} enable the construction of Android-specific call graphs that are necessary to perform inter-procedural analysis.

Validating the \ORDER section requires solving the typestate check $\sat^o_{\automatons}$. To achieve this, we use \IDEAL, a framework for efficient inter-procedural data-flow analysis~\cite{oopsla17ideal}, to instantiate a typestate analysis. The analysis defines the finite state machine $\usagePattern^o$ to check against and the allocation sites to start the analysis from.
From those allocation sites, \IDEAL performs a flow-, field-, and context-sensitive data-flow analysis. 

The constraints and the predicates require knowledge about objects and values associated with rule variables at given execution points in the program.
The typestate analysis in \TOOLSA extracts the primitive values and objects on-the-fly, where the latter are abstracted by allocation sites. 
When the typestate analysis encounters a call site that is referred to in an event definition, and the respective rule requires the object or value of an argument to the call, \TOOLSA triggers an on-the-fly backward analysis to extract the objects or values that may participate in the call. This on-the-fly analysis yields comparatively high performance and scalability, because many of the arguments of interest are values of type \code{String} and \code{Integer}. Thus, using an on-demand computation avoids constant propagation of \emph{all} strings and integers through the program. For the on-the-fly backward analysis, we extended the on-demand pointer analysis Boomerang~\cite{DBLP:conf/ecoop/SpathDAB16} to propagate both allocation sites and primitive values.

Once the typestate analysis is completed, and all required queries to Boomerang are computed, \TOOLSA solves the internal constraints and predicates using our own custom-made solvers.

%\js{@SK: I think we should not speak about the GUI and CLI here. RQ4 might be a good place to address it}
%\eb{I agree that it feels a bit out of place.} \todo{KA: briefly mention that as part of the experimental setup. It's not a core part of the implementation}

%We have implemented our analysis in two modes, GUI and command line. In command-line mode, the analysis strives to signal as many programming mistakes as possible within a single analysis run. The analysis thus evaluates the internal constraints regardless even when the typestate analysis signals a mistake already. It also keeps checking objects that require predicates from objects whose analyses already resulted in a warning.
%The GUI mode seeks to ease an integration of the analysis into an IDE. For this reason, to not overwhelm the user with warning messages and to increase the responsiveness of the analysis, it skips the evaluation of the internal constraints of an object if its corresponding typestate analysis detected an error. Objects that require a predicate from a faulty object are also not analysed themselves. Instead, one error message is generated for each such object alerting the developer to the unfulfilled predicate.
\section{Evaluation}\label{sec:evaluation}
We evaluate our implementation of \DSL and \TOOLSA by addressing the following research questions: 
\begin{enumerate}[label=\RQ{\arabic*:}]
\item What are the precision and recall of \TOOLSA? 
\item What types of misuses does \TOOLSA find?
\item How fast does \TOOLSA return results?
\item How does \TOOLSA compare to the state-of-the-art? 
\end{enumerate}

To answer these questions, we developed \DSL rules for all JCA classes. We then applied \TOOLSA using this ruleset to statically analyze \empirical{10,001} Android apps from the AndroZoo dataset~\cite{AllixBKT16}. We chose apps that are available in the official Google Play Store and received an update in 2017. This ensures that we report on the most up-to-date usages of \crapis. Our project web page lists all apps in our dataset and our \DSL ruleset to facilitate reproduction: \url{http://cryptoapis.wordpress.com}

During our evaluation, \TOOLSA frequently reported misuses within packages for commonly used libraries across different apps. To avoid over-counting the same misuses, we excluded the following common library packages: \code{com.google}, \code{com.unity3d}, \code{com.facebook.ads}, and \code{com.android}.

\subsection{Precision and Recall (\RQ{1})}
\label{sec:manual}
\subsubsection*{Setup}
To compute precision and recall, two authors of this work manually checked \empirical{50} randomly selected apps from our dataset for typestate errors and violations of internal constraints. We did not check for unsatisfied predicates or forbidden events because these are hard to detect manually. We compare the results of our manual analysis to those reported by \TOOLSA. Our goal here is to compute precision and recall of the analysis implementation in \TOOLSA, not the quality of our \DSL rules. We discuss the latter in \secref{sec:rq4}. Consequently, we define a false positive to be a warning that should not be reported according to the specified rule, irrespective of that rule's semantic correctness (similarly for false negatives).

\subsubsection*{Results}
In the \empirical{50} apps we inspected, \TOOLSA detects \empirical{228} usages of JCA classes. Table \ref{tab:manualanalysis} lists the misuses it finds. Overall, the analysis finds \empirical{156} misuses. In particular, it issues \empirical{27} typestate-related warnings, with only \empirical{2} false positives, because the analysis is path insensitive (\secref{sec:static-analysis}). We further found \empirical{4} false negatives, which are caused by initializing a \code{MessageDigest} or a \code{MAC} object without completing the operation. \TOOLSA fails to find these typestate errors, because the supporting alias analysis times out, and \TOOLSA aborts the typestate analysis without reporting a warning.

The automated analysis finds \empirical{129} violations of internal constraints. We were able to confirm \empirical{110} of them. For the other \empirical{19} cases, the analysis fails to statically extract possible runtime values for certain variables due to obfuscated code. For such values, the constraint solver reports the corresponding constraint as violated. We have also checked the apps for missed constraint violations, but were not able to locate any. 

\roundbox{\RQ{1}: Our manual analysis shows that our typestate analysis achieves high precision (\empirical{92.6\%}) and recall (\empirical{86.2\%}). The constraint resolution has a precision of \empirical{85.3\%} and a recall of \empirical{100\%}.}

\begin{table}
  \centering
    \caption{Correctness of \TOOLSA warnings.}
    \label{tab:manualanalysis}%
    \begin{tabular}{lrrr}
    \toprule
          & \multicolumn{1}{l}{\textbf{Total Warnings}} & \multicolumn{1}{l}{\textbf{False Pos.}} & \multicolumn{1}{l}{\textbf{False Neg.}} \\ \midrule
    Typestate & 27    & 2     & 4 \\
    Constraints & 129   & 19    & 0 \\ \midrule
    Total & 156   & 21    & 4 \\
    \bottomrule
    \end{tabular}%
\end{table}%

\subsection{Types of Misuses (\RQ{2})}
\subsubsection*{Setup} 
We report the results of analyzing all \empirical{10,001} Android apps from AndroZoo. We then use the results of our manual analysis (\secref{sec:manual}) as a baseline to evaluate our findings on a large scale.

\subsubsection*{Results}
\TOOLSA detects the usage of at least one JCA class in \empirical{4,071} apps (41\% of the analyzed apps). Most of these apps (\empirical{96\%}) contain at least one misuse. In total, \TOOLSA discovers \empirical{19,756} individual object traces that contradict the specified rule patterns. We categorize these misuses into the following: typestate errors~(\empirical{2,669}), unsatisfied predicates~(\empirical{3523}), forbidden events~(\empirical{159}), and internal constraint violations~(\empirical{11,436}).

The violations of internal constraints represent the largest class of misuses. Approximately \empirical{82\%} of these violations are related to \code{MessageDigest}. In our manual analysis, most violations (\empirical{89/110}) originate from usages of \code{MD5} and \code{SHA-1}. Many developers still use these algorithms, although both are no longer recommended by security experts~\cite{BSI}. \TOOLSA identifies \empirical{1,766} (\empirical{15.4\%}) constraint violations related to \code{Cipher} usages. Our manual analysis confirms that all misuses of the \code{Cipher} class are due to using the insecure algorithm \code{DES} or mode of operation \code{ECB}. This result is in line with the findings of prior studies~\cite{EgeleBFK13,ShaoDGYS14,Chatzikonstantinou:2016}. 

More than \empirical{75\%} of the typestate errors are caused by misuses of \code{MessageDigest}. Through our manual analysis, we attribute this high number to incorrect usages of \code{reset}. In addition to misusing \code{MessageDigest}, misuses of \code{Cipher} contribute \empirical{421} typestate errors. Finally, \TOOLSA detects \empirical{89} typestate errors related to \code{PBEKeySpec}. The \ORDER section of the \DSL rule for \code{PBEKeySpec} requires calling \code{clearPassword} at the end of the lifetime of a \code{PBEKeySpec} object. We manually inspected \empirical{3} of the reported misuses and observed that the invocation of \code{clearPassword} is missing in all of them.

Predicates are unsatisfied when \TOOLSA expects the interaction of multiple object traces but is not able to prove their correct interaction. With \empirical{3,523} unsatisfied predicates reported, the number may seem relatively large because unsatisfied predicates accumulate transitively. For example, if \TOOLSA cannot ensure a predicate for a usage of \code{IVParameterSpec}, it will not generate a predicate for the key object that \code{KeyGenerator} generates using the \code{IVParameterSpec} object. Transitively, \TOOLSA reports an unsatisfied predicate for a \code{Cipher} object that relies on the generated key object.

\TOOLSA also finds \empirical{159} calls to forbidden methods. As only two JCA classes require the definition of forbidden methods in our \DSL ruleset (\code{PBEKeySpec} and \code{Cipher}), we do not find this low number surprising. A manual analysis of a handful of reports suggests that most of the reported forbidden methods originate from the insecure \code{PBEKeySpec} constructors (\secref{sec:lang}).

From the \empirical{4,071} apps that use at least one JCA Crypto API, \empirical{1,757} contain at least one typestate error (\empirical{43\%}), \empirical{1,079} lack required predicates (\empirical{26.5\%}), \empirical{155} use at least one forbidden method (\empirical{3.8\%}), and \empirical{4,001} violate at least one internal constraint (\empirical{93.7\%}). Ignoring the class \code{MessageDigest}, \empirical{1,119} apps still violate at least one constraint in other classes.

\roundbox{\RQ{2}: \empirical{96\%} of apps misuse at least one Crypto API. Violating the constraints of \code{MessageDigest} is the most common type of misuse.}

\subsection{Performance (\RQ{3})}
\subsubsection*{Setup}
\TOOLSA comprises four main phases. It constructs (1) a \emph{call graph} using FlowDroid~\cite{ArztRFBBKTOM14} and then runs the actual analysis (\secref{sec:static-analysis}), which (2) calls the \emph{typestate analysis} and (3) \emph{constraint analysis} as required, attempting to (4) \emph{resolve all declared predicates}. During the analysis of our dataset, we measured the execution time that \TOOLSA spent in each phase. We ran \TOOLSA once per application and capped the time of each run to \empirical{30}~minutes.

\subsubsection*{Results}
Overall, \TOOLSA times out after 30 minutes for only \empirical{275} of all \empirical{10,001} apps in our dataset (\empirical{2.75\%}). Unfortunately, \TOOLSA crashed during the analysis of \empirical{604} apps in different phases. \figref{fig:performance} summarizes the distribution of analysis times (in seconds) for the four phases as well as the total analysis time. The numbers are reported across the remaining \empirical{9,122} apps for which the analysis successfully terminates in the allotted 30 minutes. For each phase, the box plot highlights the median, the 25\% and 75\% quartiles, and the minimal and maximal values of the distribution.
\begin{figure}
\begin{tikzpicture}[xshift=1.6cm,yshift=-1cm]
\begin{axis}
    [
    height=4cm,
    width=6.8cm,
    xlabel={Analysis Time (s)}, 
    xmode=log,
    axis line style={gray},
    ytick={1,2,3,4, 5},
    yticklabels={ Constraints, Typestate, Predicate, Call Graph, Total Time},
    yticklabel style={text width=3cm,align=center},
  boxplot/every box/.style={draw=none,fill=black},
     boxplot/every whisker/.style={gray},
      boxplot/every median/.style={white,thick},    ]

    \addplot+[
    boxplot prepared={
      lower whisker=0.001,
      lower quartile=0.33,
      median=0.64,
      upper quartile=42.2,
      upper whisker=1473.8,
    },
    ] coordinates {};
   
	\addplot+[
    boxplot prepared={
      lower whisker=0.012,
      lower quartile=1.48,
      median=3.23,
      upper quartile=32.2,
      upper whisker=1342.8,
    },
    ] coordinates {}; 
    
        \addplot+[
    boxplot prepared={
      lower whisker=0.005,
      lower quartile=0.33,
      median=0.68,
      upper quartile=1.6,
      upper whisker=421.1,
    },
    ] coordinates {};
    
    \addplot+[
    boxplot prepared={
      lower whisker=9.56,
      lower quartile=63.3,
      median=83.9,
      upper quartile=121.3,
      upper whisker=1759.3,
    },
    ] coordinates {};    
    
		\addplot+[
    boxplot prepared={
      lower whisker=9.97,
      lower quartile=70.8,
      median=108.4,
      upper quartile=230.6,
      upper whisker=1789.139,
    },
    ] coordinates {};    
    
  \end{axis}
  \end{tikzpicture}
  \caption{Performance of \TOOLSA.}
\label{fig:performance}
\end{figure}

Across the apps in our dataset, there is a very large variation in the reported execution time (between 10 seconds and 29.9 minutes). We attribute this to two main reasons. First, apps have different sizes---reachable methods in the call graph vary between \empirical{141} and \empirical{30,259} (median: \empirical{3,075}~methods). The majority of the total analysis time is spent on call-graph construction. Resolving all declared predicates takes approximately \empirical{0.6}~seconds for half of the apps, with the typestate analysis having a median runtime of \empirical{3.2}~seconds. For more than half  of the apps, the value extraction and constraint resolution finishes in less than \empirical{1} second. 

\roundbox{\RQ{3}: On average, \TOOLSA analyzes an app in \empirical{108}~seconds, with call-graph construction taking most of the time (\empirical{76\%}).}

\subsection{Comparison to Existing Tools (\RQ{4})}
\label{sec:rq4}
\subsubsection*{Setup}

We compare \TOOLSA to \cryptolint~\cite{EgeleBFK13}, the most closely related tool. Unfortunately, we were unable to obtain access to \cryptolint's implementation, despite contacting the authors. However, we were able to use \DSL to reimplement the original ruleset of \cryptolint. \DSL has generally proven expressive enough to model the \cryptolint rules, proving it is a useful specification language beyond the scope of this work.

Our original \DSL ruleset covers all JCA classes. \cryptolint, however, comprises only six individual rules. For easier distinction, we refer to our full ruleset for all JCA classes as \SARuleSet, the original rules \cryptolint uses as \CLRules, and our \DSL version for them as \CLDSLRuleSet. Both \CLRules and \CLDSLRuleSet are available at our project website. \CLRules does not include any typestate properties or forbidden methods, and hence, can be modelled using only internal constraints and predicates in \DSL. For three out of the six rules in \CLRules, \DSL expresses exactly the checks that \cryptolint performs. The remaining three rules (3,4, and 6 in~\cite{EgeleBFK13}) cannot be directly expressed. \cryptolint rule~4, for instance, requires non-constant values for salts in \code{PBEKeySpec}. In \DSL such a relationship is expressed through predicates. However, predicates model correct behaviour only. Therefore, in \DSL we had to further strengthen this \cryptolint rule: we created a rule for \code{PBEKeySpec} that requires the salt to be random. We followed a similar approach with the other two rules in \CLRules. Despite being more strict than \CLRules, \CLDSLRuleSet ensures a fair comparison between \TOOLSA and \cryptolint: when comparing the two tools in terms of their findings, the stricter rules in \CLDSLRuleSet tend to produce more warnings than \CLRules, which works in favor of \cryptolint.

\subsubsection*{Results}
%\SARuleSet
%\CLRules
%\CLDSLRuleSet

Using \CLDSLRuleSet, \TOOLSA detects usages of JCA classes in \empirical{1,726} Android apps. In total, it reports \empirical{6,098} misuses, only a third of roughly \empirical{20,000} misuses that \TOOLSA identifies using the \SARuleSet. For each of the four types of misuses, \TOOLSA finds more apps using \SARuleSet. Using \CLDSLRuleSet, all reported warnings are related to \empirical{6} classes, compared to \empirical{14} using \SARuleSet. The differences mainly stem from three types of misuses. As we have pointed out, \CLDSLRuleSet does not specify any typestate properties or forbidden methods. Hence, it does not find approximately \empirical{3,000} warnings that \TOOLSA identifies in these categories using \SARuleSet. Furthermore, while \TOOLSA reports \empirical{11,436} constraint violations using \SARuleSet, it reports only \empirical{1,356} using \CLDSLRuleSet.

To our surprise, significantly fewer apps violate four of the six original rules in \CLRules. For example, for \cryptolint rule~1 that forbids the use of \code{ECB} mode for encryption, \cryptolint identified \empirical{7,656} apps breaking this rule (\empirical{65.2\%} of apps that use \crapis). Using \CLDSLRuleSet, \TOOLSA identifies \empirical{658} usages of ECB mode in \empirical{38.1\%} of apps that use \crapis. Although a high number of apps still exhibit this basic misuse, there is a considerable decrease compared to previous studies.

\roundbox{\RQ{4}: The more comprehensive \TOOLSA ruleset detects \empirical{3$\times$} as many misuses as \cryptolint in \empirical{twice} as many JCA classes.}

\subsection{Threats to Validity}
Our ruleset is mainly based on the documentation of the JCA~\cite{JCArefguide2017}. 
Although the authors of this paper have significant domain expertise, our \DSL-rule specifications for the JCA are only as correct as the JCA documentation. Our static analysis toolchain depends on multiple external components. Yet, of course, we cannot fully rule out bugs in the implementation.

Java allows a developer to programmatically select a non-default cryptographic service provider. \TOOLSA currently does not detect such customizations but instead assumes that the default provider is used. This behaviour may lead to imprecise results, because our rules forbid certain default values that are insecure for the default provider but may be secure for a different one.

\section{Related Work}

In this section, we discuss languages that specify API properties and tools that detect misuses of security APIs.

\paragraph*{Specifying API Properties}

There is a significant body of research on textual specification languages that ensure API properties by means of static data-flow analysis. For example, tracematches~\cite{AllanACHKLMSST05} enable runtime-checking typestate properties defined by regular expressions over runtime objects. Bodden et at.~\cite{ICSE10Cont,toplas2012} as well as \citet{NaeemL08} present algorithms to (partially) evaluate state matches prior to the program execution, using static analysis.

\citet{MartinLL05} present Program Query Language (PQL) that enables a developer to specify patterns of event sequences that constitute potentially defective behaviour. A combination of static and dynamic analyses match the patterns to a given program. A pattern may include a fix that is applied to each match by dynamic instrumentation. PQL has been applied to detecting security-related vulnerabilities such as memory leaks~\cite{MartinLL05}, SQL injection and cross-site scripting~\cite{LivshitsL05}. Compared to tracematches, PQL captures a greater variety of pattern specifications, at the disadvantage of using a flow-insensitive static analysis. PQL serves as the main inspiration for the \DSL syntax. Other languages that pursue similar goals include PTQL~\cite{GoldsmithOA05}, PDL~\cite{MorganVW07}, and TS4J~\cite{Bodden14}.

These languages and their analysis-tool support are different from \DSL and \TOOLSA in three main aspects. First, these systems follow a black-list approach by defining and finding incorrect program behaviour. On the other hand, \DSL rules define desired behaviour, which in the case of \crapis leads to more compact specifications. Second, the above languages are general-purpose languages for bug finding, while \DSL specifically targets misuses of \crapis, which may seem a limitation of \DSL. However, the stronger focus on cryptography allows us to cover a greater portion of cryptography-related problems in \DSL compared to other languages, while at the same time keeping \DSL relatively simple. Third, \TOOLSA uses state-of-the-art static analyses that have superior performance and precision compared to other static-analysis approaches~\cite{oopsla17ideal}.

\paragraph*{Detecting Misuses of Security APIs}

Throughout the paper, we have discussed \cryptolint~\cite{EgeleBFK13}, and compared it to \TOOLSA in \secref{sec:evaluation}. Another tool that finds misuses of \crapis is Crypto Misuse Analyzer (CMA)~\cite{ShaoDGYS14}. The CMA ruleset has significant overlaps with our \DSL ruleset. However, the CMA rules are limited to misuses related to encryption and hashing. Unlike \TOOLSA, CMA has been evaluated on a small dataset of only 45 apps. \citet{Chatzikonstantinou:2016} ran a dynamic checker for a number of misuses and manually verified their findings on 49 apps. All three studies concluded that at least \empirical{88\%} of the studied apps misuse at least one Crypto~API.

Unlike \TOOLSA, none of these tools facilitates rule creation by means of a higher-level specification language. Instead, the rules are hard-coded into the tool, making it hard for non-experts to extend or alter the ruleset. Due to its Java-like syntax, \DSL enables regular developers---including cryptography experts---to define their own rules. \TOOLSA then automatically transforms those rules into the appropriate static analysis checks. Finally, \TOOLSA includes a typestate analysis that checks for generally forbidden methods. 

%When translated to \DSL, all their rules would either be turned into internal constraints and predicates. In this sense, the tools currently have orthogonal expressiveness, which means that it may make sense to use them in combination.

%This approach also has drawbacks, though. As one design goal of \DSL was ease of use, we aimed to not overload the language with new elements and introduce new ones with care. This limits \DSL 's expressiveness. For instance, \citet{ShaoDGYS14}'s CMA correctly identifies as a misuse the use of the same cryptographic key for multiple purposes (e.g., once for encryption, once to create a MAC). At this point, this property cannot be expressed in \DSL. 

%- \citet{GeorgievIJABS12}: SSL certificate validation, misused by major apps\\
%- \citet{FahlHMSBF12}: SSL API misuse, many apps vulnerable to person-in-the-middle-attacks \\

%- \citet{HeRCCVYZ15}: \\
\section{Conclusion}\label{sec:conclusion}
In this paper, we present \DSL, a description language for correct usages of cryptographic APIs. Each \DSL rule is specific to one class, and it may include usage pattern definitions and constraints on parameter values. Predicates model the interactions between classes. For example, a rule may generate a predicate on an object if it is used successfully, and another rule may require that predicate from an object that it uses. We also present \TOOLSA, a static analyzer that checks a given program for its compliance with our \DSL ruleset. Applying \TOOLSA to \empirical{10,001} Android apps, we found \empirical{20,000} misuses spread over \empirical{96\%} of the \empirical{4,071} apps that use the JCA. \TOOLSA terminates successfully in under \empirical{2}~minutes for more than half of the apps.

In future work, we plan to address the following challenges.
% EB: I have removed the following for now, as I do not want to open an attack surface for now, where reviewers start to question whether our rules are actually correct.
% First, we have written all the rules used in \TOOLSA. While we have acquired some deeper familiarity with cryptographic concepts, we are not cryptographers ourselves. Therefore, we will reach out to cryptography experts to correct potential mistakes in our existing rules. We would also encourage domain experts to model their own cryptographic libraries in \DSL to improve the support in \TOOLSA. 
\DSL currently only supports a binary understanding of security---a usage is either secure or not. We would like to enhance \DSL to have a more fine-grained notion of security. This notion will allow for more nuanced warnings in \TOOLSA. This is challenging, because the \DSL language still ought to be concise.
Also, \DSL currently requires one rule per class per JCA provider, because there is no way to express the commonality and variability between different providers implementing the same algorithms. This leads to specification overhead. To address this, we plan to modularize the language using import and override mechanisms. Moreover, we plan to consider extending \DSL to support more complex properties such as using the same cryptographic key for multiple purposes. Finally, we plan to improve the performance of \TOOLSA through incremental static analysis~\cite{Arzt:2014:REU:2568225.2568243}.

\section*{Acknowledgments}
This work was supported by the DFG through its Collaborative Research Center CROSSING and the project RUNSECURE, by the Natural Sciences and Engineering Research Council of Canada, by the Heinz Nixdorf Foundation, a Fraunhofer ATTRACT grant and by an Oracle Collaborative Research Award. We would also like to thank the maintainers of Androzoo for allowing us to use their data set in our evaluation.
%\eb{TODO: Oracle Grant}

\bibliographystyle{ACM-Reference-Format}
\bibliography{bibfiles/references}

\end{document}